# Affine Dihedral Subgroup of $W_a(B_n)$, the symmetry of the lattice $\mathbb{Z}^n$ and Quasicrystallography

Nazife Ozdes Koca [a*], Mehmet Koca [b], Ramazan Koc [c] and Amira Al-Maqbali [a]

[a]Department of Physics, College of Science, Sultan Qaboos University, P.O. Box 36, Al-Khoud 123, Muscat, Sultanate of Oman, *Correspondence e-mail: nazife@squ.edu.om
[b]Department of Physics, Cukurova University, Adana, Turkey, retired
[c]Department of Physics, Gaziantep University, Gaziantep, Turkey

Quasicrystals described as the projections of higher dimensional cubic lattices, and the particular affine extension $W_a(I_2(h))$ of the dihedral group $W(I_2(h))$ of order $2h$, $h=2n$ being the Coxeter number, as a subgroup of affine group $W_a(B_n)$ offers a different perspective to $h$-fold symmetric quasicrystallography. The affine group $W_a(I_2(h))$ is constructed as the subgroup of the affine group $W_a(B_n)$, the symmetry of the cubic lattice $\mathbb{Z}^n$. The infinite discrete group with local dihedral symmetry of order $2h$ operates on the concentric $h$-gons obtained by projecting the Voronoi cell of the cubic lattice with $2^n$ vertices onto the Coxeter plane. Voronoi cells tile the space facet to facet, consequently, leading to the tilings of the Coxeter plane with some overlaps of the rhombic tiles. It is noted that the projected Voronoi cell is the overlap of $h$ copies of the $h$-gons tiled with some rhombi and rotated by the angle $\frac{2\pi}{h}$. After a general discussion on the lattice $\mathbb{Z}^n$ with its affine group $W_a(B_n)$ embedding the affine dihedral group $W_a(I_2(h))$ as a subgroup, its projection onto the Coxeter plane has been worked out with some examples. The cubic lattices with affine symmetry $W_a(B_n)$, $(n=1,2,3,4,5)$ have been presented and shown that the projection of the lattice $\mathbb{Z}^3$ leads to the hexagonal lattice, the projection of the lattice $\mathbb{Z}^4$ describes the Amman-Beenker quasicrystal lattice with 8-fold local symmetry and the projection of the lattice $\mathbb{Z}^5$ describes a quasicrystal structure with local 10-fold symmetry with thick and thin rhombi. It is then straight forward to show that the projections of the cubic lattices with even higher dimensions onto the Coxeter plane may lead to the quasicrystal structures with 12-fold, 18-fold symmetries and so on.

## 1. Introduction

Projections of the higher dimensional lattices onto the plane with the cut-and-project technique lead to the quasicrystallographic structures. For a general review see the reference (Senechal, 1995). Planar quasicrystallography displays local $h$-fold symmetries (Di Vincenzo & Steinhardt, 1991) and (Janot, 2012). The cut-and-project technique is far from able to explain the emergence of the local $h$-fold symmetries. In this paper we explain how the local $h$-fold symmetries arise from the projection of the higher dimensional lattices tessellated by the Voronoi cell. Here $h$ is the Coxeter number of the Coxeter-Weyl group. Symmetry of the Voronoi cell is the point Coxeter-Weyl group which has a maximal dihedral subgroup of order $2h$ (Steinberg, 1951; Carter, 1987). Affine extension of the Coxeter-Weyl group embeds the affine dihedral group. We will illustrate how the affine dihedral group as the subgroup of the affine Coxeter-Weyl group explains the rhombic tilings of the plane with local $h$-fold symmetries.

Voronoi cells of the lattices described by the Coxeter-Weyl groups (Conway & Sloane, 1988, 1991) tessellate the $n$-dimensional Euclidean space facet to facet. Facets of the Voronoi cells of some important lattices have been worked out earlier (Koca et. al., 2012, Koca et. al., 2014a, Koca et. al., 2018). Affine Coxeter-Weyl groups (Humphreys, 1990) represent the symmetries of the associated lattices and describe the tessellations of the Euclidean space by their Voronoi cells. Many of these groups admit non-crystallographic subgroups such as the Coxeter groups $W(H_2)$, $W(H_3)$ and $W(H_4)$ can be embedded into the Coxeter-Weyl groups $W(A_4)$, $W(D_6)$ and $W(E_8)$ respectively. Formal affine extensions of the groups $W(H_2)$, $W(H_3)$ and $W(H_4)$ similar to the affine extensions of the Coxeter-Weyl groups were first discussed in the reference (Patera & Twarock, 2002) and possible other affine extensions have been discussed in the reference (Dechant et.al., 2012). Affine extension of $W(H_2)$ as a subgroup of the affine Coxeter-Weyl group $W(A_4)$ is used to demonstrate the 5-fold symmetric quasicrystallography by projecting the root and weight lattices of $W(A_4)$ onto the Coxeter plane (Koca et. al., 2022). The Coxeter-Weyl group $W(D_6)$ embeds the icosahedral group $W(H_3)$ as a subgroup (Kramer, 1993) and can be used as the tilings of the 3D-Euclidean space with four composite tiles (Mosseri & Sadoc, 1982) displaying local dodecahedral structures (Koca et. al., 2021). Quasicrystallographic local $W(H_4)$ symmetry has been studied in (Elser & Sloane, 1987) and the embedding of $W(H_4)$ into $W(E_8)$ has been worked out by Coxeter (Coxeter, 1973) and its quaternionic embedding is studied in the reference (Koca et al, 2001). It seems that the affine extension of $W(H_3)$ may play important roles not only in quasicrystals but also in the viral structures (Keef & Twarock, 2008; Indelicato et. al., 2012; Dechant et al, 2013; Salthouse et. al., 2015; Zappa et. al., 2016).

De Bruijn (de Bruijn, 1981) proved that the Penrose rhombic tiling with 5-fold local symmetry can be obtained by the projection of the 5D cubic lattice onto a Euclidean plane embedded into the 4D space of $W(A_4)$ and later it has been extended to the projections of higher dimensional cubic lattices (Sire et al., 1989; Whittaker & Whittaker, 1987; Koca et.al. 2015). Earlier (Koca et. al., 2014a), by employing the cut-and-project technique, it has been shown that the quasicrystallographic tessellation of the plane can be obtained by the projections of the Delone cells of the quaternionic root lattice of $W(A_4)$. In most of the work above, the cut-and- project technique have been employed except the one in the reference (Koca et. al. 2022) where affine subgroup $W_a(I_2(5))$ of affine $W_a(A_4)$ is introduced as part of the rhombic tiling of the Coxeter plane. Affine dihedral subgroups of the affine Coxeter-Weyl groups display different features as we will work out in what follows. Extended Coxeter diagram corresponding to the group $W_a(I_2(h))$ and its embedding into the affine Coxeter-Weyl group is totally different depending on whether $h$ is odd or even. This is a different topic of research which will be discussed elsewhere (under preparation). The group $W_a(I_2(h))$ as a subgroup of the group $W_a(B_n)$ for $h = 2n$ requires a detailed study with its application on the quasicrystallography.

It is well known that the local symmetries of the 2D quasicrystallography can be described by the dihedral group $W(I_2(h))$. But up until now, neither their affine extensions have been constructed by using the extended Coxeter graphs nor the affine dihedral groups as subgroups of the affine Coxeter-Weyl groups have been studied. In what follows we determine the affine dihedral subgroup $W_a(I_2(h))$ of the affine Coxeter-Weyl group $W_a(B_n)$ and study its role in the tilings of the Euclidean plane. We will illustrate our work by exemplifying some low dimensional cubic lattices. The paper is organized as follows.

In Section 2 we discuss the problem in its generality by introducing the affine Coxeter-Weyl group $W_a(B_n)$ using the extended Coxeter-Dynkin diagram and its affine dihedral subgroup $W_a(I_2(h))$. We determine the diagrammatic extension of the point group $W(I_2(h))$, the symmetry of the projected Voronoi cell of the lattice $\mathbb{Z}^n$, and give its reflection elements as the products of the generators of the Coxeter-Weyl group $W(B_n)$. By block-diagonalization of the generators of $W(I_2(h))$ we determine the principal planes of the cubic lattice $\mathbb{Z}^n$. Section 3 introduces first three simpler cases, $W_a(B_1)$ the symmetry of the lattice $\mathbb{Z}$, $W_a(B_2) \cong W_a(I_2(4))$ the symmetry of the square lattice, and then work with $W_a(B_3)$, the symmetry of the simple cubic lattice and its projection onto the Coxeter plane leading to the hexagonal lattice described by the affine dihedral group $W_a(I_2(6))$. After this warm-up section we discuss in Section 4 the projection of the 4D cubic lattice $\mathbb{Z}^4$ by using the affine subgroup $W_a(I_2(8))$ of $W_a(B_4)$ leading to the quasicrystal lattice tiled with the squares and rhombi displaying the local 8-fold symmetries. The Section 5 deals with the similar problem of projection of the $\mathbb{Z}^5$ lattice and the role of the affine dihedral group $W_a(I_2(10))$ leading to the 10-fold symmetric tiling of the Coxeter plane with Penrose rhombs. In Section 6 we discuss our technique comparing the other projection techniques and give hints on application of the same technique to obtain the 12-fold and the 18-fold symmetric quasicrystallographic lattices. The relevant lattices are the $\mathbb{Z}^6$ and $\mathbb{Z}^9$ with the respective affine dihedral subgroups $W_a(I_2(12))$ and $W_a(I_2(18))$ of the affine groups $W_a(B_6)$(Koca et. al., 2014b), and $W_a(B_9)$.

## 2. Affine $W_a(B_n)$ and its affine dihedral subgroup $W_a(I_2(h))$

Before we proceed further, we will introduce the notations used throughout the text. The Coxeter-Dynkin diagram for the Coxeter-Weyl group $W(B_n)$ is denoted by $B_n$ and its extended diagram by $\tilde{B}_n$. The group generated by the extended diagram is the affine group $W_a(B_n)$. Similarly, the Coxeter diagram for the dihedral group of order $2h$ and its extended diagram are denoted by $I_2(h)$ and $\tilde{I}_2(h)$ respectively. Its point group and its affine group are represented respectively by $W(I_2(h))$ and $W_a(I_2(h))$. The point symmetry of the unit cell of the cubic lattice $\mathbb{Z}^n$ is defined by the Coxeter-Dynkin diagram $B_n$ as shown in Fig. 1.

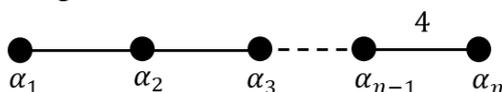

**Figure 1**
The Coxeter-Dynkin diagram of $W(B_n)$.

The nodes are used either to denote the simple roots $\alpha_i = l_i - l_{i+1}$, ($i = 1,2,\ldots,n-1$) and $\alpha_n = l_n$ where $l_i$ are orthonormal vectors of the $n$-dimensional Euclidean space satisfying $(l_i, l_j) = \delta_{ij}$ or to denote the reflection generators $r_{\alpha_i}$ acting on an arbitrary vector $\lambda$ as

$$r_{\alpha_i}(\lambda) = \lambda - \frac{2(\lambda, \alpha_i)\alpha_i}{(\alpha_i, \alpha_i)}. \tag{1}$$

The set of roots $\pm l_i \pm l_j$ are called the long roots and the roots $\pm l_i$ are the short roots of $B_n$. The lattice $\mathbb{Z}^n$ is generated by the short roots and a general vector of the lattice is given by $\sum_i k_i l_i$, $k_i \in \mathbb{Z}$. One of the interesting subgroups of the group $W(B_n)$ is the dihedral group $W(I_2(h))$ generated by two generators $R_1$ and $R_2$ defined as (Steinberg, 1951; Carter, 1972)

$$R_1 = r_{\alpha_1} r_{\alpha_3} \ldots r_{\alpha_n}, R_2 = r_{\alpha_2} r_{\alpha_4} \ldots r_{\alpha_{n-1}} \text{ for odd } n, \qquad (2a)$$
$$R_1 = r_{\alpha_1} r_{\alpha_3} \ldots r_{\alpha_{n-1}}, R_2 = r_{\alpha_2} r_{\alpha_4} \ldots r_{\alpha_n} \text{ for even } n. \qquad (2b)$$

In either case the Coxeter element $R = R_1 R_2$ satisfies $R^h = 1$ where the Coxeter number is $h = 2n$ and the dihedral group is represented by the diagram $I_2(h)$ as given in Fig. 2. The generators $R_1$ and $R_2$ are $n \times n$ matrices in the $l_i$ basis with integer entries.

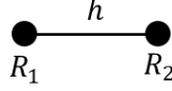

**Figure 2**
Diagrammatic representation of the dihedral group $W(I_2(h))$ with its generators.

The extended Coxeter-Dynkin diagram $\tilde{B}_n$ leading to the affine Coxeter-Weyl group $W_a(B_n)$ is depicted in Fig. 3 where the extended node is the root $\alpha_0 = -(l_1 + l_2)$.

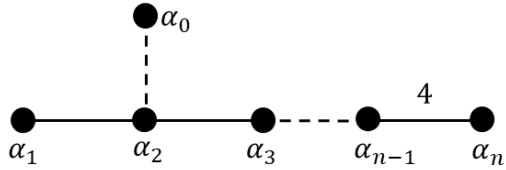

**Figure 3**
Extended Coxeter-Dynkin diagram $\tilde{B}_n$.

The affine Coxeter-Weyl group $W_a(B_n)$ is generated by the reflections $r_{\alpha_i}$, ($i = 1, 2, \ldots, n-1$), $r_{\alpha_n}$ and $r_{\alpha_0,1}$ where the affine reflection for an arbitrary root and integer $k$ is denoted by $r_{\alpha,k}$ and defined by the action on an arbitrary vector $\lambda$ (Humphreys, 1990) as

$$r_{\alpha,k}(\lambda) = \lambda - \frac{2((\lambda,\alpha) - k)\alpha}{(\alpha,\alpha)}. \qquad (3)$$

For some detailed properties of the affine generators $r_{\alpha,k}$ one can consult Appendix A of the reference (Koca et. al, 2022).

Extended Coxeter diagram $\tilde{I}_2(h)$ of the dihedral subgroup of the Coxeter-Weyl group $W_a(B_n)$ is illustrated in Fig. 4 which has not been worked out elsewhere.

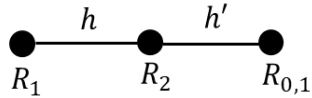

**Figure 4**
Extended Coxeter diagram $\tilde{I}_2(h)$.

In Fig. 4, $h' = \frac{2h}{h-2} = \frac{2n}{n-1}$ is obtained by letting the determinant of the affine extension of the Cartan matrix

$$\det \begin{bmatrix} 2 & -2\cos\frac{\pi}{h} & 0 \\ -2\cos\frac{\pi}{h} & 2 & -2\cos\frac{\pi}{h} \\ 0 & -2\cos\frac{\pi}{h} & 2 \end{bmatrix} = 0 \qquad (4)$$

and the extended node represents the generator $R_{0,1}$ given by

$$R_{0,1} = (R_1 R_2)^n R_1. \tag{5}$$

In Fig. 5 some of the diagrams of the extended dihedral groups are shown.

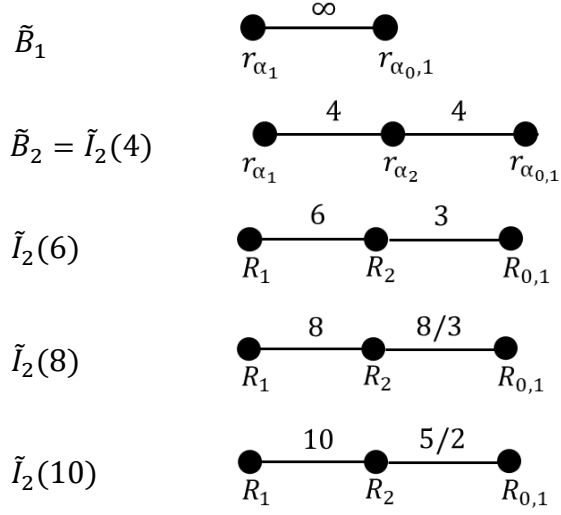

**Figure 5**
Extended diagrams of some affine dihedral groups $W_a(I_2(h))$ as subgroups of affine $W_a(B_n)$ where the number, $m$, between the nodes represents the angle $\pi/m$ between the intersecting reflection line segments.

The principal planes (Engel, 1986) of the cubic lattice $\mathbb{Z}^n$ are determined by block-diagonalization of the matrices $R_1$ and $R_2$ with a choice of a new set of orthonormal vectors given by (Koca et. al, 2015)

$$\hat{x}_i = \frac{1}{\sqrt{h\lambda_i}} \sum_j \frac{2\alpha_j}{(\alpha_j, \alpha_j)} X_{ji}, \tag{6}$$

where $X_{ji}$ is the eigenvector $\vec{X}_i$ of the Cartan matrix of $W(B_n)$ corresponding to the eigenvalue

$$\lambda_i = 2\left[1 + \cos\left(\frac{m_i \pi}{h}\right)\right]. \tag{7}$$

The normalization of the eigenvectors is such that the last components all equal 1. Here $m_i$ are the Coxeter exponents of $W(B_n)$ taking the values $m_i = 1, 3, 5, \ldots, 2n-1$. The pairs of unit vectors $(\hat{x}_i, \hat{x}_{n+1-i})$ determine the principal planes, one of which specified by the pair of vectors $(\hat{x}_1, \hat{x}_n)$ is known as the Coxeter plane. The simple roots of the dihedral group $I_2\left(\frac{h}{m_i}\right)$ can be determined as (Koca et. al., 2015)

$$\beta_i = \sqrt{2}\left[\sin\left(\frac{m_i \pi}{2h}\right)\hat{x}_i + \cos\left(\frac{m_i \pi}{2h}\right)\hat{x}_{n+1-i}\right],$$
$$\beta_{n+1-i} = \sqrt{2}\left[\sin\left(\frac{m_i \pi}{2h}\right)\hat{x}_i - \cos\left(\frac{m_i \pi}{2h}\right)\hat{x}_{n+1-i}\right], \left(i = 1, 2, \ldots, \frac{n}{2}\right). \tag{8}$$

The orbits $W(B_n)\omega_i$ of the weight vectors $\omega_i$ under the group $W(B_n)$ determine certain polytopes (Koca et. al., 2015). For example, the orbit of the weight $\omega_1 = l_1$ consists of the vectors $\pm l_i$ forming an $n$-dimensional octahedron or the cross polytope. The orbit of the weight $\omega_n$ consisting of $2^n$ vectors of the form $\frac{1}{2}(\pm l_1 \pm l_2 \pm \cdots \pm l_n)$ is an $n$-dimensional cube determining the vertices of the Voronoi cell $V(0)$ which can be taken as the unit cell of the dual lattice implying that the lattice $\mathbb{Z}^n$ is self-dual. The facets of the Voronoi cell $V(0)$ are the $(n-1)$-dimensional cubes whose centers are half the unit vectors $\pm \frac{l_i}{2}$. The unit cells of the root lattice $\mathbb{Z}^n$ can be considered as the Delone cells where the Voronoi cells centralize the Delone cells. In terms of the coordinates of the Voronoi cells the cubic lattice can be represented by the vectors $\left(k_1 + \frac{1}{2}\right)l_1 + \left(k_2 + \frac{1}{2}\right)l_2 + \cdots + \left(k_n + \frac{1}{2}\right)l_n$. In the rest of the paper, we will be working with the lattice $\mathbb{Z}^n$ tiled by the Voronoi cell. Orthogonal projection of the lattice onto the Coxeter plane will be obtained by taking the components of the vectors $l_i$ in the $(\hat{x}_1, \hat{x}_n)$ plane. This preliminary introduction on the lattice $\mathbb{Z}^n$ is to understand its projection onto the Coxeter plane via Voronoi cells and the action of the affine $W_a(I_2(h))$ on the rhombic tiles leading to the quasicrystallographic lattices for $n \geq 4$. For an arbitrary root $\alpha$ and integer $k$ the affine reflection formula for the *lattice generated by the Voronoi cell* is defined as

$$r_{\alpha,k}(\lambda) = \lambda - \frac{2((\lambda,\alpha))\alpha}{(\alpha,\alpha)} + k\alpha, \quad k \in \mathbb{Z}. \tag{9}$$

Equation (3) is replaced by (9) as it applies on the lattice generated by the Voronoi cell with the vectors $\left(k_1 + \frac{1}{2}\right)l_1 + \left(k_2 + \frac{1}{2}\right)l_2 + \cdots + \left(k_n + \frac{1}{2}\right)l_n$, which also represents (3) as a special case for $k$ is even in (9). Before we study the affine group, we first discuss the point symmetry of the Voronoi cell $V(0)$.

**Projection of the Voronoi cell $V(0)$**

Vertices of the Voronoi cell $V(0)$ decompose into the orbits of concentric $h$-gons after projection onto the Coxeter plane. With a suitable choice of the simple roots of $W(B_n)$ we find a Coxeter element $R = R_1 R_2$ which permutes the $2n$ unit vectors in a cyclic order projected on to the Coxeter plane,

$$R: l_1 \to l_2 \to \cdots \to l_n \to -l_1 \to -l_2 \to \cdots \to -l_n. \tag{10}$$

These are the vertices of the $n$-octahedron, after projection, forming an $h$-gon in the Coxeter plane with $2n$ vertices and $2n$ edges. The symmetry of the $h$-gon is the point dihedral group $W(I_2(h))$ with $h$ rotation and $h$ reflection elements. The reflection elements $R_1, R_2$ and $(R_1 R_2)^i R_1$, ($i = 1, 2, \ldots, h-2$) represent the reflections with respect to the diagonals and the lines bisecting the edges of an $h$-gon. The rotation elements $R^i (i = 1, 2, \ldots, h)$ permute cyclically the vertices of the $h$-gons. While some vertices of the Voronoi cell project into the concentric $h$-gons and some other vertices project directly into the origin. Table 1 lists the projections of some Voronoi cells of the cubic lattices which have not been studied elsewhere. Entries in the third column represent the decomposition of the vertices of the Voronoi cell $V(0)$ as the number of orbits of the dihedral group $W(I_2(h))$ forming $h$-gons. The 4th column gives the number of vertices of $V(0)$ projected into the origin. The 5th column denotes the type of rhombs formed by projection of the faces of the Voronoi cell $V(0)$. Since the angle between the projected adjacent pairs of vectors is $\frac{2\pi}{h}$ the square faces of the Voronoi cells project as rhombs with angles of integer multiples of $\frac{2\pi}{h}$. The fact that the Voronoi cell tiles the $n$-dimensional Euclidean space facet to facet it is expected that the projected quasi-lattice will display a tiling with overlaying rhombs. Not only the lattice but also each projected Voronoi cell display a tiling with overlaying rhombs. As we will show in the remaining sections that the projected Voronoi cell can be dissociated into concentric $h$-gons each rotated by $\frac{2\pi}{h}$ with respect to each other and each one is tiled with nonoverlapping rhombs. Anyone of this concentric $h$-gon tiled by nonoverlapping rhombs possesses only one reflection symmetry line of the dihedral group $W(I_2(h))$ so that when they overlap, they constitute the projected Voronoi cell invariant under the group $W(I_2(h))$. Since each set of concentric $h$-gon tiled by rhombs is identical up to a rotation we can work only with one of them and apply the affine dihedral group to generate a tiling scheme of the plane with rhombs.

**Table 1**. Projection of the Voronoi cell of the lattice $\mathbb{Z}^n$ onto the Coxeter plane.

| $n$ | # of vertices of $V(0)$ | # of concentric polygons | # of vertices projecting into the origin | Faces project into rhombs with angles |
|---|---|---|---|---|
| 1 | 2 | line-segment | none | - |
| 2 | 4 | square | none | $\left(\frac{\pi}{2},\frac{\pi}{2}\right)$ |
| 3 | 8 | 1 hexagon | 2 | $\left(\frac{\pi}{3},\frac{2\pi}{3}\right)$ |
| 4 | 16 | 2 octagons | none | $\left(\frac{\pi}{2},\frac{\pi}{2}\right), \left(\frac{\pi}{4},\frac{3\pi}{4}\right)$ |
| 5 | 32 | 3 decagons | 2 | $\left(\frac{\pi}{5},\frac{4\pi}{5}\right), \left(\frac{2\pi}{5},\frac{3\pi}{5}\right)$ |
| 6 | 64 | 5 dodecagons | 4 | $\left(\frac{\pi}{6},\frac{5\pi}{6}\right), \left(\frac{\pi}{3},\frac{2\pi}{3}\right), \left(\frac{\pi}{2},\frac{\pi}{2}\right)$ |
| 7 | 128 | 9, 14-gons | 2 | $\left(\frac{\pi}{7},\frac{6\pi}{7}\right), \left(\frac{2\pi}{7},\frac{5\pi}{7}\right), \left(\frac{3\pi}{7},\frac{4\pi}{7}\right)$ |
| 8 | 256 | 16, 16-gons | none | $\left(\frac{\pi}{8},\frac{7\pi}{8}\right), \left(\frac{\pi}{4},\frac{3\pi}{4}\right), \left(\frac{3\pi}{8},\frac{5\pi}{8}\right), \left(\frac{\pi}{2},\frac{\pi}{2}\right)$ |
| 9 | 512 | 28, 18-gons | 8 | $\left(\frac{\pi}{9},\frac{8\pi}{9}\right), \left(\frac{2\pi}{9},\frac{7\pi}{9}\right), \left(\frac{\pi}{3},\frac{2\pi}{3}\right), \left(\frac{4\pi}{9},\frac{5\pi}{9}\right)$ |
| 10 | 1024 | 51, 20-gons | 4 | $\left(\frac{\pi}{10},\frac{9\pi}{10}\right), \left(\frac{\pi}{5},\frac{4\pi}{5}\right), \left(\frac{3\pi}{10},\frac{7\pi}{10}\right), \left(\frac{2\pi}{5},\frac{3\pi}{5}\right), \left(\frac{\pi}{2},\frac{\pi}{2}\right)$ |

**Matrix Representation of the Affine Group $W_a(B_n)$**

A general affine transformation $(g, \lambda)$ of the lattice $\mathbb{Z}^n$ where $g$ is an element of the group $W(B_n)$ fixing the origin and the vector $\lambda$ representing the translation can be written as a $(n+1) \times (n+1)$ matrix in the form

$$\begin{bmatrix} g_{11} & g_{12} & \cdots & g_{1n} & \lambda_{\hat{x}_1} \\ g_{21} & g_{22} & \cdots & g_{2n} & \lambda_{\hat{x}_2} \\ \vdots & \vdots & \vdots & \ddots & \vdots \\ g_{n1} & g_{n2} & \cdots & g_{nn} & \lambda_{\hat{x}_n} \\ 0 & 0 & 0 & 0 & 1 \end{bmatrix}. \qquad (11)$$

Here $g = [g_{ij}]$ is a $n \times n$ matrix, $\lambda = (\lambda_{\hat{x}_1}, \lambda_{\hat{x}_2}, \ldots, \lambda_{\hat{x}_n})^T$ and a vector $x \in R^n$ is represented by a column $(x_1, x_2, \ldots, x_n, 1)^T$.

It is clear that the product of two group elements satisfies the relation $(g_1, \lambda_1)(g_2, \lambda_2) = (g_1 g_2, g_1 \lambda_2 + \lambda_1)$ where the unit element and the inverse are given by $(I, 0)$ and $(g, \lambda)^{-1} = (g^{-1}, -g^{-1}\lambda)$ respectively.

Let us assume that the element of the point group represents one of the generators of $W(I_2(h))$ along with a translation vector. After block-diagonalization of the generators, that is, transforming the matrix into sets of $2 \times 2$ matrices and, $1 \times 1$ matrix in the case of odd $n$, the affine dihedral group operating in the Coxeter plane is represented by $3 \times 3$ matrices with the addition of a 2-component translation vector. Details of this work can be found in the forthcoming sections where we discuss the Coxeter-Weyl group $W_a(B_n)$ for $(n = 1,2,3,4,5)$ and its affine dihedral subgroup and the projection of the lattice onto the Coxeter plane. Only in the case of $(n = 4,5,...)$ we obtain quasicrystallographic tilings with rhombs.

## 3. Affine $W_a(B_n)$ for $(n = 1,2,3)$ and their affine dihedral subgroups $W_a(I_2(h))$

### 3.1. The affine group $W_a(B_1)$ as the symmetry of the integer lattice $\mathbb{Z}$

Here the point group is generated by a single generator $r_{\alpha_1}$ with $\alpha_1 = l_1$ and the generator corresponding to its affine extension is given by the generator $r_{\alpha_1,1}$ which can be represented by the $2 \times 2$ matrices as follows

$$r_{\alpha_1} = \begin{bmatrix} -1 & 0 \\ 0 & 1 \end{bmatrix}, \quad r_{\alpha_1,1} = \begin{bmatrix} -1 & 1 \\ 0 & 1 \end{bmatrix}, \tag{12}$$

where the diagrammatic representation is given in Fig. 5. They are the generators of the affine group $W_a(B_1)$ with infinite number of elements representing the symmetry of the integer lattice where the unit cell is the line segment between (0 and 1) (Delone cell) and the Voronoi cell centralizing the lattice point (0) is the line segment between the points $\pm \frac{1}{2}$ so the lattice generated by the Voronoi cell consists of the infinite set of half integers. Facets of the Voronoi cell $V(0)$ are the points $\pm \frac{1}{2}$. Reflection with respect to the point $\frac{1}{2}$ is represented by the operator $r_{\alpha_1,1}$ and the reflection with respect to the point $-\frac{1}{2}$ is represented by the operator $r_{-\alpha_1,1} = r_{\alpha_1,-1}$. The product of two reflections, say, $r_{\alpha_1,1} r_{\alpha_1}$ represents the translation by one unit in the positive direction. The affine group is the infinite dihedral group representing the symmetry of an apeirogon, an infinite straight line broken into equal segments.

### 3.2. The affine $W_a(B_2)$ representing the symmetry of the square lattice $\mathbb{Z}^2$

The simple roots of the point group are $\alpha_1 = l_1 - l_2, \alpha_2 = l_2$ and the matrix representations of the generators of the affine group $W_a(B_2)$ are given by

$$r_{l_1-l_2} = \begin{bmatrix} 0 & 1 & 0 \\ 1 & 0 & 0 \\ 0 & 0 & 1 \end{bmatrix}, \quad r_{l_2} = \begin{bmatrix} 1 & 0 & 0 \\ 0 & -1 & 0 \\ 0 & 0 & 1 \end{bmatrix}, \quad r_{-l_1-l_2,1} = \begin{bmatrix} 0 & -1 & -1 \\ -1 & 0 & -1 \\ 0 & 0 & 1 \end{bmatrix}. \tag{13}$$

The affine Coxeter-Dynkin diagram is shown in Fig. 5. The roots consist of the vectors $(\pm l_1 \pm l_2), \pm l_i$. The root lattice consists of the vectors $m_1 l_1 + m_2 l_2, m_i \in \mathbb{Z}$. The lattice generated by the Voronoi cell consists of the vectors $\left(m_1 + \frac{1}{2}\right) l_1 + \left(m_2 + \frac{1}{2}\right) l_2$. The product $r_{-l_1-l_2} r_{-l_1-l_2,1}$ is a translation along $-(l_1 + l_2)$. The square lattice and its Voronoi cell $V(0)$ with the vertices $\left(\pm \frac{1}{2}, \pm \frac{1}{2}\right)$ are shown in Fig. 6.

The Coxeter element $r_{l_1-l_2} r_{l_2}$ is a rotation of order 4 which rotates the lattice around the origin but $r_{l_1-l_2} r_{l_2,1}$ is a 4-fold rotation around the point $\left(\frac{1}{2}, \frac{1}{2}\right)$ by using equation (9). So, one can obtain the dihedral group fixing either one of the lattice point or one of the vertices of the Voronoi cell. The facets of the Voronoi cell $V(0)$ are the edges of the square bisected by the short roots $\pm l_i$. In this example the point group $W(B_2)$ is itself a dihedral group of order 8. The reflection elements of the point groups represent the reflections with respect to the diagonals of the Voronoi cell $V(0)$ as well as reflections with respect to the line segments bisecting the edges. The reflections with respect to the edges of the Voronoi cell $V(0)$ are represented by the generators $r_{l_i,\pm 1}$. The reflections with respect to the lines passing through the vertices of the Voronoi cell $V(0)$ and parallel to the diagonals are represented by the operators $r_{\pm l_1 \pm l_2, 1}$. These are some examples as to how the elements of the group $W_a(B_2)$ operate on the square lattice.

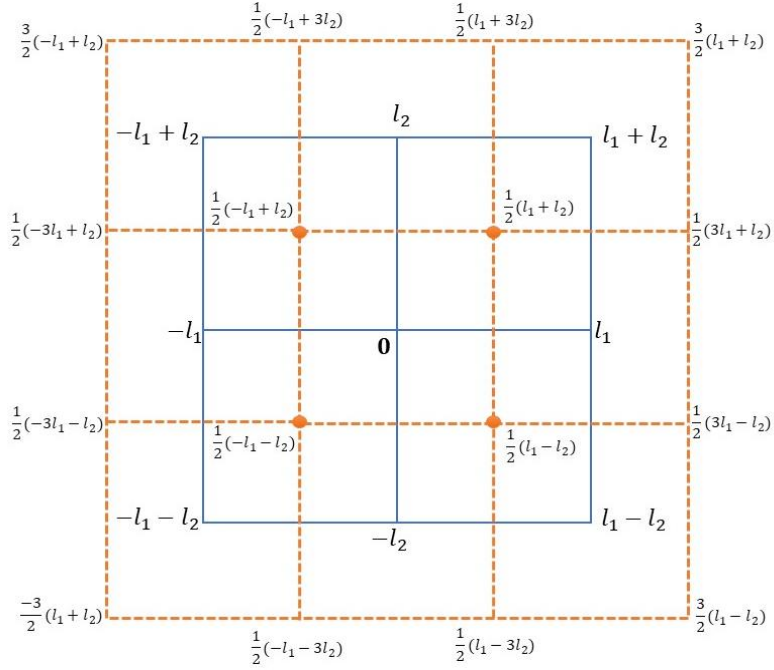

**Figure 6**
Square lattice of roots and the lattice generated by its Voronoi cell $V(0)$.

### 3.3. The affine group $W_a(B_3)$ as the symmetry of the cubic lattice $\mathbb{Z}^3$

In this subsection we will also discuss the projection of the cubic lattice generated by the Voronoi cell $V(0)$. The choice of a basis of the simple roots is as many as the order of the Coxeter-Weyl group and a suitable basis can be chosen for a simpler representation of the Coxeter element $R = R_1 R_2$. Let the simple roots be chosen as $\alpha_1 = l_2 - l_3$, $\alpha_2 = l_3 - l_1$ and $\alpha_3 = l_1$. The generators of the dihedral subgroup will be $R_1 = r_{l_1} r_{l_2 - l_3}$ and $R_2 = r_{l_3 - l_1}$ where the Coxeter element of the group $R = R_1 R_2$ permutes the unit vectors as $R: l_1 \to l_2 \to l_3 \to -l_1 \to -l_2 \to -l_3$. The affine generators of the group $W_a(B_3)$ can be written in the basis $l_i$ as $4 \times 4$ matrices:

$$r_{l_2-l_3} = \begin{bmatrix} 1 & 0 & 0 & 0 \\ 0 & 0 & 1 & 0 \\ 0 & 1 & 0 & 0 \\ 0 & 0 & 0 & 1 \end{bmatrix}, r_{l_3-l_1} = \begin{bmatrix} 0 & 0 & 1 & 0 \\ 0 & 1 & 0 & 0 \\ 1 & 0 & 0 & 0 \\ 0 & 0 & 0 & 1 \end{bmatrix}, r_{l_1} = \begin{bmatrix} -1 & 0 & 0 & 0 \\ 0 & 1 & 0 & 0 \\ 0 & 0 & 1 & 0 \\ 0 & 0 & 0 & 1 \end{bmatrix}, \qquad (14a)$$

and the affine generator

$$r_{-l_2-l_3,1} = \begin{bmatrix} 1 & 0 & 0 & 0 \\ 0 & 0 & -1 & -1 \\ 0 & -1 & 0 & -1 \\ 0 & 0 & 0 & 1 \end{bmatrix}. \qquad (14b)$$

The matrix generators in (14a) generate the octahedral group of order 48 leaving the origin invariant and with the addition of the generator of (14b) the group is extended to the infinite group $W_a(B_3)$ which is the symmetry group of the simple cubic (SC) lattice $\mathbb{Z}^3$. Of course, the simple cubic lattice is a well-known lattice but here we study it with the Coxeter-Dynkin diagram technique. The affine group can also be generated by the affine reflections defined by $r_{l_2-l_3,n_2}$, $r_{l_3-l_1,n_3}$ and $r_{l_1,n_1}$ representing the symmetry of the lattice generated by the unit cell $V(\lambda)$. Using equation (9) for each simple root we can find the vector $\lambda$ fixed by the affine transformations. For the fixed values of $n_1, n_2, n_3$ these generators generate the octahedral group leaving the lattice vector

$$\lambda = \tfrac{n_1}{2} l_1 + \left(\tfrac{n_1}{2} + n_2 + n_3\right) l_2 + \left(\tfrac{n_1}{2} + n_3\right) l_3, \tag{15}$$

invariant. The vector in (15) is a vertex of the cube. For even $n_1$ this is the lattice generated by the vectors $l_i$ with integer coefficients. It is clear that for $n_1 = n_2 = n_3 = 0$ this is the octahedral group leaving the origin invariant. For odd $n_1$, $\lambda$ is a vector of the lattice generated by the Voronoi cell $V(0)$. Facets of the Voronoi cells $V(0)$ are the square faces whose centers are represented by the vectors $\pm \tfrac{l_i}{2}$ and the generators $r_{l_i,\pm 1}$ represent the reflections with respect to these faces of the Voronoi cell $V(0)$. The product of the generators $r_{-l_2-l_3}r_{-l_2-l_3,1}$ is a translation by the vector $l_2 + l_3$. If we consider the generators pairwise, for example, the generators $(r_{l_2-l_3,n_2}, r_{l_3-l_1,n_3})$ generate a triangular lattice while the generators $(r_{l_3-l_1,n_3}, r_{l_1,n_1})$ generate a square lattice and the pair of generators $(r_{l_2-l_3,n_2}, r_{l_1,n_1})$ the lattice $\mathbb{Z} \times \mathbb{Z}$.

Our main interest here, of course, is the affine dihedral group generated by $R_1 = r_{l_1,n_1} r_{l_2-l_3,n_2}$ and $R_2 = r_{l_3-l_1,n_3}$ operating on the Coxeter plane and the projection of the lattice onto the Coxeter plane. With the use of the formula (6) we determine the components of the vectors $l_i$ in terms of the unit vectors $\hat{x}_i$ and choosing the Coxeter plane as defined by the unit vectors $\hat{x}_1$ and $\hat{x}_3$ the generators of the dihedral group $W(I_2(6))$ of order 12 can be transformed into block diagonal forms as $2 \times 2$ and $1 \times 1$ matrices as shown below

$$R_1' = \begin{bmatrix} -\frac{\sqrt{3}}{2} & -\frac{1}{2} & 0 \\ -\frac{1}{2} & \frac{\sqrt{3}}{2} & 0 \\ 0 & 0 & -1 \end{bmatrix}, R_2' = \begin{bmatrix} -\frac{\sqrt{3}}{2} & \frac{1}{2} & 0 \\ \frac{1}{2} & \frac{\sqrt{3}}{2} & 0 \\ 0 & 0 & 1 \end{bmatrix}. \tag{16}$$

One can take only the upper $2 \times 2$ matrices as they operate only on the first $\hat{x}_1$ and $\hat{x}_3$ components of the vectors $l_i$. The extended Coxeter-Dynkin diagram of $W(I_2(6))$ is shown in Fig. 5 with the generators $R_i'$ where $i = 0, 1, 2$. The generator $R_0' = -R_1'$ represents the extended node in Fig. 5. Obviously, we have the relations $(R_1' R_2')^6 = 1$ and $(R_2' R_0')^3 = 1$.

The Coxeter plane cuts the Voronoi cell at the origin orthogonal to the diagonal of the cube through the vertices $\pm \tfrac{1}{2}(l_1 - l_2 + l_3)$ so that the projected components satisfy the relation $l_{p2} = l_{p1} + l_{p3}$. The lattice is generated by the vectors $l_{p1}, l_{p3}$ where a general element of the lattice is obtained from (15) as

$$\lambda_p = k_1 l_{p1} + k_3 l_{p3}, \quad k_1, k_3 \in \mathbb{Z}, \quad (l_{pi}, l_{pi}) = \tfrac{2}{3}. \tag{17}$$

The angle between $l_{p1}$ and $l_{p3}$ is $120^0$. The six reflection elements of the group $W(I_2(6))$ generated by $R_1'$ and $R_2'$ represent the reflections with respect to the diagonals and the bisectors of the edges of the hexagon and the remaining 6 elements represent the rotations. Affine dihedral group possesses the elements fixing the centers and the vertices of the hexagons so that the centers and the vertices of the hexagons are the hexagonal lattice points. One vertex is at the center of the other hexagon satisfying the 6-fold symmetry with respect to both the center of the hexagon as well as with respect to any one of the vertices. The tiles of the lattice are equilateral triangles as shown in Fig. 7.

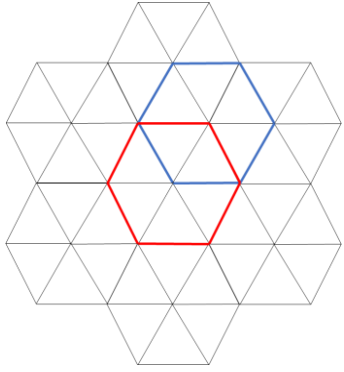

**Figure 7**
Hexagonal lattice obtained from projection of $\mathbb{Z}^3$ onto the Coxeter plane. The red and blue hexagons imply that the hexagonal lattice is not only invariant under the dihedral group $W(I_2(6))$ fixing the centers of the hexagons but also fixing the vertices of a given hexagon.

## 4. Affine $W_a(B_4)$ as the symmetry of $\mathbb{Z}^4$ and its projection with 8-fold symmetry

So far, we have summarized the properties of the lattices $\mathbb{Z}, \mathbb{Z}^2$ and $\mathbb{Z}^3$ and their symmetries. We have also obtained the hexagonal lattice as the projection of the cubic lattice described by the affine dihedral group with a point symmetry $W(I_2(6))$ of order 12. We have shown that the projections of the lattice $\mathbb{Z}^3$ onto four different planes lead to the lattices defined by the symmetries $W_a(B_1) \times W_a(B_1)$, $W_a(A_2)$ with triangular symmetry, $W_a(B_2)$ with square symmetry and the $W_a(I_2(6))$ with hexagonal symmetry. All projections lead to the lattice structures with 2-fold, 3-fold, 4-fold and 6-fold lattice symmetries. However, as we will illustrate the projections of the lattices $\mathbb{Z}^n$ for $n \geq 4$ onto the Coxeter plane render the quasicrystal structures with affine dihedral symmetries $W_a(I_2(2n))$ with $2n$-fold local symmetries.

Although the projection of the cubic lattice $\mathbb{Z}^4$ has been studied earlier (Whittaker & Whittaker, 1987; Koca et.al. 2015) the importance of the affine dihedral subgroup $W_a(I_2(h))$ for $n \geq 4$ of the affine group $W_a(B_n)$ has not been emphasized and has not been studied in the form of tilings by rhombs. In this section we will illustrate the tiling of the Coxeter plane by squares and rhombs of angles of $45^0$ and explain why the affine dihedral group $W_a(I_2(8))$ leads to the desired 8-fold symmetric tessellation of the plane.

Since we work with the lattice generated by the Voronoi cell the lattice vectors are given by

$$\left(k_1 + \frac{1}{2}\right) l_1 + \left(k_2 + \frac{1}{2}\right) l_2 + \left(k_3 + \frac{1}{2}\right) l_3 + \left(k_4 + \frac{1}{2}\right) l_4, \ k_i \in \mathbb{Z}, \tag{18}$$

where the components of the unit vectors $l_i$ are obtained from (6) in the order of unit vectors $\hat{x}_1, \hat{x}_4, \hat{x}_2, \hat{x}_3$ as

$$l_1 = \tfrac{1}{\sqrt{2}}\left[\cos\left(\tfrac{13\pi}{16}\right), \sin\left(\tfrac{13\pi}{16}\right), \cos\left(\tfrac{7\pi}{16}\right), \sin\left(\tfrac{7\pi}{16}\right)\right]^T,$$
$$l_2 = \tfrac{1}{\sqrt{2}}\left[\cos\left(\tfrac{9\pi}{16}\right), \sin\left(\tfrac{9\pi}{16}\right), \cos\left(\tfrac{27\pi}{16}\right), \sin\left(\tfrac{27\pi}{16}\right)\right]^T,$$
$$l_3 = \tfrac{1}{\sqrt{2}}\left[\cos\left(\tfrac{5\pi}{16}\right), \sin\left(\tfrac{5\pi}{16}\right), \cos\left(\tfrac{15\pi}{16}\right), \sin\left(\tfrac{15\pi}{16}\right)\right]^T,$$
$$l_4 = \tfrac{1}{\sqrt{2}}\left[\cos\left(\tfrac{\pi}{16}\right), \sin\left(\tfrac{\pi}{16}\right), \cos\left(\tfrac{3\pi}{16}\right), \sin\left(\tfrac{3\pi}{16}\right)\right]^T. \tag{19}$$

The simple roots of $W(B_4)$ in (6) were taken as

$$\alpha_1 = l_2 - l_3, \; \alpha_2 = l_3 - l_1, \; \alpha_3 = l_1 - l_4 \text{ and } \alpha_4 = l_4. \tag{20}$$

First two components of the vector $l_i$ define the Coxeter plane and the last two components are in the orthogonal plane so that the generators $R_1 = r_{l_2-l_3} r_{l_1-l_4}, R_2 = r_{l_3-l_1} r_{l_4}$ of the dihedral group $W(I_2(8))$ are converted into the block-diagonal forms:

$$R'_1 = \begin{bmatrix} -\cos\left(\frac{\pi}{8}\right) & \sin\left(\frac{\pi}{8}\right) & 0 & 0 \\ \sin\left(\frac{\pi}{8}\right) & \cos\left(\frac{\pi}{8}\right) & 0 & 0 \\ 0 & 0 & -\cos\left(\frac{3\pi}{8}\right) & \sin\left(\frac{3\pi}{8}\right) \\ 0 & 0 & \sin\left(\frac{3\pi}{8}\right) & \cos\left(\frac{3\pi}{8}\right) \end{bmatrix},$$

$$R'_2 = \begin{bmatrix} -\cos\left(\frac{\pi}{8}\right) & -\sin\left(\frac{\pi}{8}\right) & 0 & 0 \\ -\sin\left(\frac{\pi}{8}\right) & \cos\left(\frac{\pi}{8}\right) & 0 & 0 \\ 0 & 0 & -\cos\left(\frac{3\pi}{8}\right) & -\sin\left(\frac{3\pi}{8}\right) \\ 0 & 0 & -\sin\left(\frac{3\pi}{8}\right) & \cos\left(\frac{3\pi}{8}\right) \end{bmatrix}. \tag{21}$$

Because of the above choice of the simple roots the Coxeter element $R = R_1 R_2$ permutes the vectors $l_i$ as

$$l_1 \to l_2 \to l_3 \to l_4 \to -l_1 \to -l_2 \to -l_3 \to -l_4. \tag{22}$$

The upper $2 \times 2$ matrices operate in the Coxeter plane and the 8 reflection elements of the dihedral group $R'_1, R'_2, (R'_1 R'_2)^i R'_1$, $(i = 1,2,...,6)$ represent the reflections with respect to the bisectors of the edges and the diagonals of the octagon formed by 8 vectors $\pm l_i$. We can also represent the reflection elements of the dihedral group as the products of the reflection elements of the point group $W(B_4)$ as

$$\begin{aligned} R_1 &= r_{l_2-l_3} r_{l_1-l_4}, & (R_1 R_2)^2 R_1 &= r_{l_1+l_2} r_{l_3-l_4}, \\ (R_1 R_2)^4 R_1 &= r_{l_2+l_3} r_{l_1+l_4}, & (R_1 R_2)^6 R_1 &= r_{-l_1+l_2} r_{l_3+l_4}, \end{aligned} \tag{23a}$$

$$\begin{aligned} R_2 &= r_{l_3-l_1} r_{l_4}, & (R_1 R_2) R_1 &= r_{l_4-l_2}, r_{l_1}, \\ (R_1 R_2)^3 R_1 &= r_{l_3+l_1} r_{l_2}, & (R_1 R_2)^5 R_1 &= r_{l_2+l_4} r_{l_3}. \end{aligned} \tag{23b}$$

The reflection elements in (23a) and (23b) represent the reflections with respect to the line segments joining the opposite vertices and the line segments bisecting the opposite edges of the octagon shown in Fig. 8 respectively.

Obviously the Coxeter-Weyl group $W(B_4)$ leaves the Voronoi cell $V(0)$ with vertices $\frac{1}{2}(\pm l_1 \pm l_2 \pm l_3 \pm l_4)$ invariant. Its projection onto the Coxeter plane is depicted in Fig. 8 which is invariant under the dihedral group $W(I_2(8))$.

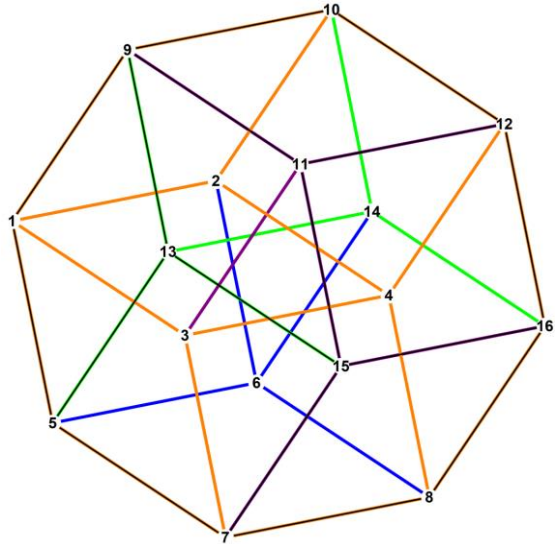

**Figure 8**
Projection of the Voronoi cell $V(0)$ of the lattice $\mathbb{Z}^4$. The numbers in the figure represent the vertices $\frac{1}{2}(\pm l_1 \pm l_2 \pm l_3 \pm l_4)$ and the colors are explained in Figure 9.

A general set of generators of the affine dihedral group $W_a(I_2(8))$ can be given as

$$R_1 = r_{l_2-l_3,n_2} r_{l_1-l_4,n_1}, \quad R_2 = r_{l_3-l_1,n_3} r_{l_4,n_4}, \quad n_i \in \mathbb{Z}. \tag{24}$$

For a fixed set of integers $n_i \in \mathbb{Z}$ the generators in (24) generate a dihedral group fixing the point $\lambda$:

$$\lambda = \left(n_3 + \frac{n_4}{2}\right) l_1 + \left(n_1 + n_2 + n_3 + \frac{n_4}{2}\right) l_2 + \left(n_2 + n_3 + \frac{n_4}{2}\right) l_3 + \frac{n_4}{2} l_4. \tag{25}$$

The vector $\lambda$ represents the root lattice for even $n_4$ and it represents the lattice generated by the Voronoi cell for odd $n_4$.

For all values of $n_i \in \mathbb{Z}$ the group is the affine extension $W_a(I_2(8))$ consisting of reflections and translations. For $n_1 = n_2 = n_3 = 0$ and $n_4 = 1$, equation (25) represents the vertex $\frac{1}{2}(l_1 + l_2 + l_3 + l_4)$, one of the vertices of the Voronoi cell $V(0)$ and the generators $R_1$ and $R_2$ generate the dihedral group leaving the vector $\frac{1}{2}(l_1 + l_2 + l_3 + l_4)$ invariant.

From now on we will be working with the projected components of the vectors $l_{pi}$ of $l_i$ which are represented by the first two components of the vectors in (19) where the scalar products of the vectors $(l_{p1}, l_{p3}) = 0$ and $(l_{p2}, l_{p4}) = 0$ hold; we will use $l_i$ for $l_{pi}$ to avoid frequent repetitions of the notation $l_{pi}$. In the Coxeter plane only two vectors are linearly independent and the following relations between the projected vectors hold

$$\begin{aligned} l_2 - l_4 = \sqrt{2} l_1, \quad l_1 + l_3 = \sqrt{2} l_2, \quad l_2 + l_4 = \sqrt{2} l_3, \quad l_3 - l_1 = \sqrt{2} l_4, \\ l_2 - l_3 = \tan\left(\frac{\pi}{8}\right)(l_1 - l_4), \quad l_1 + l_4 = \tan\left(\frac{\pi}{8}\right)(l_2 + l_3), \\ l_2 - l_1 = \tan\left(\frac{\pi}{8}\right)(l_3 + l_4), \quad l_3 - l_4 = \tan\left(\frac{\pi}{8}\right)(l_1 + l_2) \end{aligned} \tag{26}$$

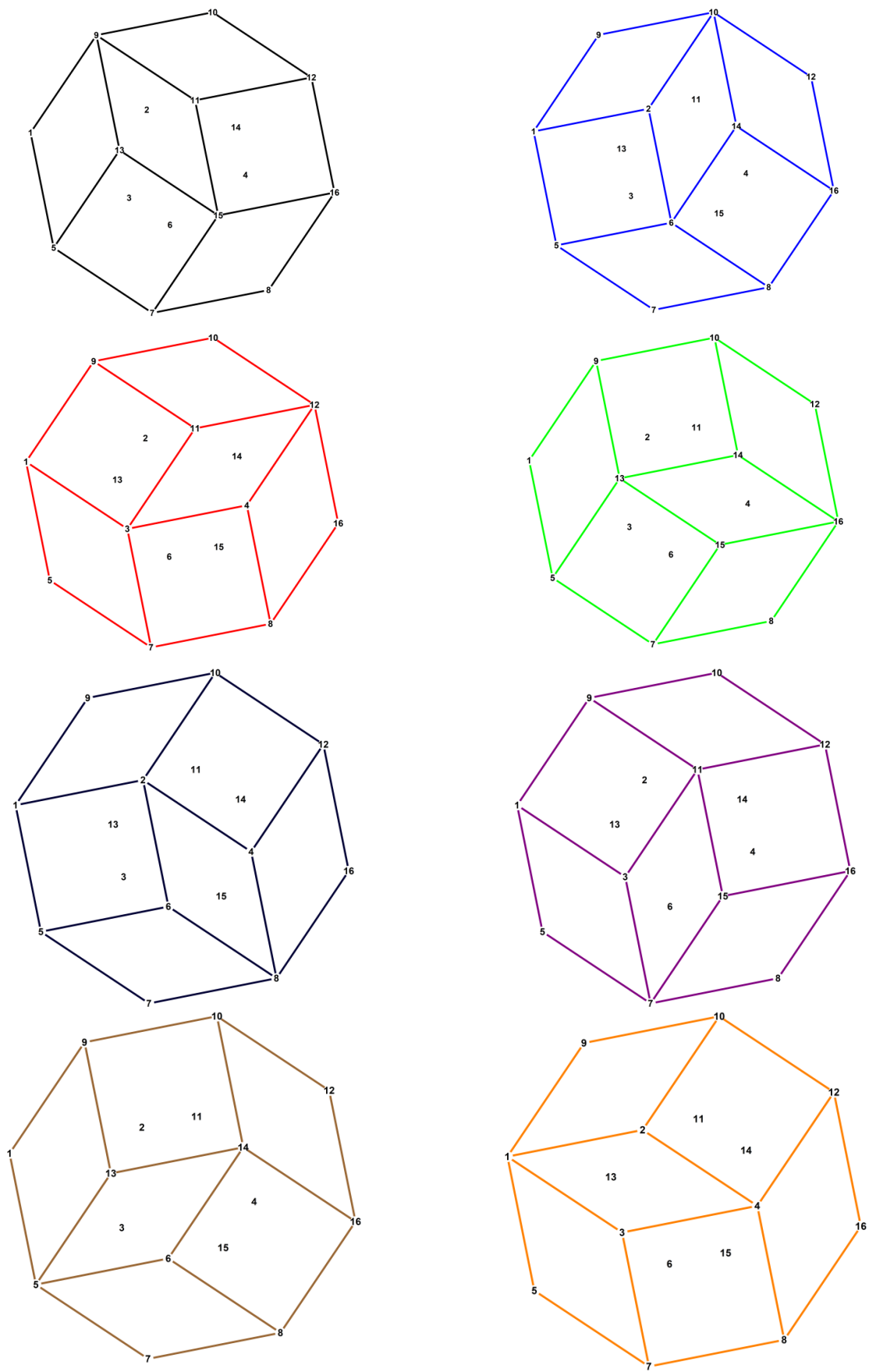

**Figure 9**
Dissociation of the projected Voronoi cell $V(0)$ of $W_a(B_4)$ depicted in Fig. 8 into 8 octagons tiled by rhombs and squares. The octagons shown in different colors are copies of each other rotated by $\frac{2\pi}{8}$.

These relations also follow from each other by cyclic permutations of the vectors $l_i$. The projected Voronoi cell $V(0)$ consists of two concentric octagons with the ratio of radii $\left(\tan\left(\frac{\pi}{8}\right)\right)^{-1}$ which can be dissociated into 8 patches of tiles, each of which, is tiled with 4 rhombi with acute angle $45^0$ and 2 squares and rotated with respect to each other by $\frac{2\pi}{8}$ as shown in Fig. 9.

The first tiling in Fig. 9 is symmetric with respect to the line segment between the vertices $\frac{1}{2}(l_1 + l_2 - l_3 - l_4)$ and $-\frac{1}{2}(l_1 + l_2 - l_3 - l_4)$. It represents the reflection by the reflection element $r_{l_2+l_3}r_{l_1+l_4}$ of the dihedral group. Similarly, the next 7 tilings of the octagon represent the symmetries with respect to the remaining reflection elements in (23). Since each one is symmetric with respect to one line of reflection of the dihedral group, when they are overlapped, they produce the projection of the Voronoi cell $V(0)$ of $W_a(B_4)$ invariant under the dihedral group $W(I_2(8))$. Note also in the first octagon in Fig. 9 we see the projection of two adjacent cubes sharing one square face. The centers of the cubes are represented by the halves of the vectors $\frac{l_2}{2}$ and $-\frac{l_3}{2}$. The other pairs of adjacent cubes can be seen in the following tilings by permuting the pairs $(l_2, -l_3)$. The line segment from $\frac{1}{2}(l_1 + l_2 - l_3 - l_4)$ to $\frac{1}{2}(-l_1 + l_2 + l_3 + l_4)$ represents the reflection by the affine reflection element $r_{l_1+l_3}r_{l_2,1}$. The angle between this line segment and the one before is $\frac{\pi}{8}$ therefore the corresponding two reflection elements generate the dihedral group fixing the vector $\frac{1}{2}(l_1 + l_2 - l_3 - l_4)$ and the product of two generators represents a rotation by $\frac{2\pi}{8}$ which produces the tiling in Fig. 10. Similar arguments valid for the other tiles. That is to say, the other tiles in Fig. 9 leads to the same tile in Fig. 10 except the centers of symmetry are obtained from the vector $\frac{1}{2}(l_1 + l_2 - l_3 - l_4)$ by cyclic rotations of the vectors $l_i$.

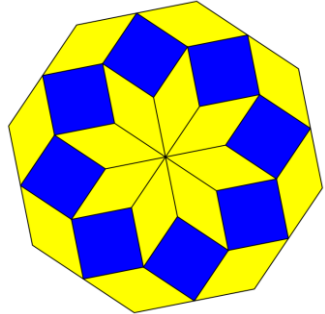

**Figure 10**
Tiling obtained by rotation of one tile in Fig. 9.

Whatever is the group operation on Fig. 10 there exists a corresponding group operation on the other 7 copies of the tilings. They are identical up to a rotation and translation. Therefore, it is sufficient to work with one copy to tessellate the plane by the rhombs by the affine group operations. Since the rhombs and squares of facets of the Voronoi cell overlap under the affine group $W_a(I_2(8))$ we display one of the dissociated patches of the tessellation in Fig. 11. Fig. 11 is obtained by translating and rotating one of the octagons in Fig. 9 with the elements of $W_a(I_2(8))$ to fill the next layer of Fig.10 so that the finite patch is invariant under the finite group $W(I_2(8))$. This process continues to increase the number of tiles in the symmetric patch.

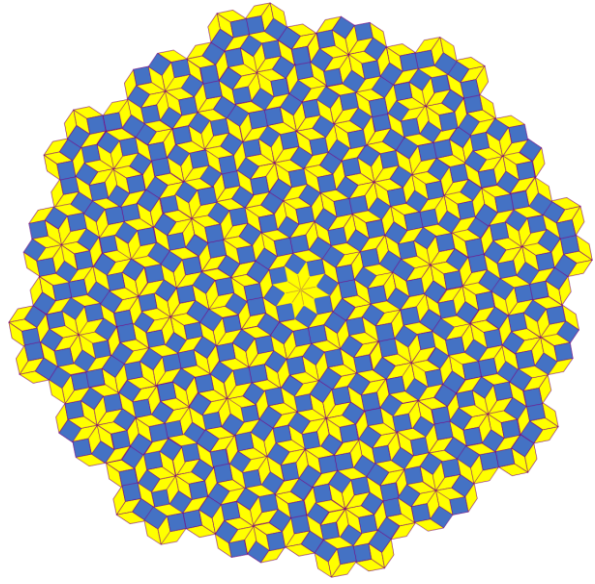

**Figure 11**
A patch of 8-fold symmetric tessellation with rhombs and squares.

## 5. Affine $W_a(B_5)$ as the symmetry of $\mathbb{Z}^5$ and its projection with 10-fold symmetry

The projection of the lattice $\mathbb{Z}^5$ has been proposed by De Bruijn (de Bruijn, 1981) to produce the Penrose tiling with 5-fold symmetry. Earlier [Koca et.al., 2014a], we have shown that the Bruijn's projection followed two steps; first projection was into the 4D space with the symmetry of $W(A_4)$ then the second projection onto a plane. The Bruijn's technique produces a tiling of the plane with 5-fold symmetry without any consideration of the dihedral subgroup. Actually, in a recent paper (Koca et al, 2022), we have obtained the Penrose tiling with the use of affine dihedral subgroup $W_a(I_2(5))$ as a subgroup of $W_a(A_4)$. In this section we obtain a tiling of the plane with thin and thick rhombs displaying 10-fold symmetry for we use the affine dihedral subgroup $W_a(I_2(10))$ of the affine Coxeter-Weyl group $W_a(B_5)$. The work is direct application of the general procedure discussed in Section 2.

The extended Coxeter-Dynkin diagram of $B_5$ is shown in Fig. 12 with its simple roots and its extended root as follows

$$\alpha_1 = l_2 - l_3, \ \alpha_2 = l_3 - l_1, \ \alpha_3 = l_1 - l_4, \ \alpha_4 = l_4 + l_5, \ \alpha_5 = -l_5, \ \alpha_0 = -l_2 - l_3. \qquad (27)$$

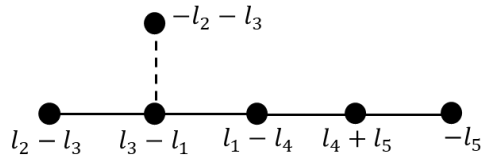

**Figure 12**
Extended diagram of $B_5$.

When the generators of the dihedral subgroup $W(I_2(10))$ is chosen as

$$R_1 := r_{l_2-l_3} r_{l_1-l_4} r_{-l_5}, \ \ R_2 := r_{l_3-l_1} r_{l_4+l_5}, \qquad (28)$$

the Coxeter element $R = R_1 R_2$ permutes the vectors $l_i$ in the cyclic order. The order of the group $W(I_2(10))$ is 20 with 10 elements representing the reflections with respect to the diagonals and the bisectors of the edges of the decagon whose vertices are represented by $\pm l_i$. In the order of the unit vectors $\hat{x}_1, \hat{x}_5, \hat{x}_2, \hat{x}_4, \hat{x}_3$ the components of the vectors $l_i$ are given by,

$$l_1 = \sqrt{\frac{2}{5}}\left[\cos\left(\frac{\pi}{4}\right), \sin\left(\frac{\pi}{4}\right), \cos\left(\frac{3\pi}{4}\right), \sin\left(\frac{3\pi}{4}\right), -\frac{1}{\sqrt{2}}\right]^T,$$

$$l_2 = \sqrt{\frac{2}{5}}\left[\cos\left(\frac{9\pi}{20}\right), \sin\left(\frac{9\pi}{20}\right), \cos\left(\frac{27\pi}{20}\right), \sin\left(\frac{27\pi}{20}\right), \frac{1}{\sqrt{2}}\right]^T,$$

$$l_3 = \sqrt{\frac{2}{5}}\left[\cos\left(\frac{13\pi}{20}\right), \sin\left(\frac{13\pi}{20}\right), \cos\left(\frac{39\pi}{20}\right), \sin\left(\frac{39\pi}{20}\right), -\frac{1}{\sqrt{2}}\right]^T,$$

$$l_4 = \sqrt{\frac{2}{5}}\left[\cos\left(\frac{17\pi}{20}\right), \sin\left(\frac{17\pi}{20}\right), \cos\left(\frac{11\pi}{20}\right), \sin\left(\frac{11\pi}{20}\right), \frac{1}{\sqrt{2}}\right]^T,$$

$$l_5 = \sqrt{\frac{2}{5}}\left[\cos\left(\frac{21\pi}{20}\right), \sin\left(\frac{21\pi}{20}\right), \cos\left(\frac{23\pi}{20}\right), \sin\left(\frac{23\pi}{20}\right), -\frac{1}{\sqrt{2}}\right]^T. \tag{29}$$

$$R'_1 = \begin{bmatrix} -\cos\left(\frac{\pi}{10}\right) & -\sin\left(\frac{\pi}{10}\right) & 0 & 0 & 0 \\ -\sin\left(\frac{\pi}{10}\right) & \cos\left(\frac{\pi}{10}\right) & 0 & 0 & 0 \\ 0 & 0 & -\cos\left(\frac{3\pi}{10}\right) & -\sin\left(\frac{3\pi}{10}\right) & 0 \\ 0 & 0 & -\sin\left(\frac{3\pi}{10}\right) & \cos\left(\frac{3\pi}{10}\right) & 0 \\ 0 & 0 & 0 & 0 & -1 \end{bmatrix},$$

$$R'_2 = \begin{bmatrix} -\cos\left(\frac{\pi}{10}\right) & \sin\left(\frac{\pi}{10}\right) & 0 & 0 & 0 \\ \sin\left(\frac{\pi}{10}\right) & \cos\left(\frac{\pi}{10}\right) & 0 & 0 & 0 \\ 0 & 0 & -\cos\left(\frac{3\pi}{10}\right) & \sin\left(\frac{3\pi}{10}\right) & 0 \\ 0 & 0 & \sin\left(\frac{3\pi}{10}\right) & \cos\left(\frac{3\pi}{10}\right) & 0 \\ 0 & 0 & 0 & 0 & 1 \end{bmatrix}. \tag{30}$$

The Coxeter element represents the rotation in the counter clockwise direction by $\frac{\pi}{5}$. Upper $2 \times 2$ matrices of the matrices $R'_1$ and $R'_2$ operate in the Coxeter plane. The Voronoi cell with 32 vertices project onto the Coxeter plane as three concentric decagons and two vertices project into the origin as shown in Fig. 13. Vertices of the three decagons, under Coxeter element $R$, are obtained by permuting the vertices $p_1 = \frac{1}{2}(l_1 + l_2 + l_3 + l_4 + l_5)$ with radius $\sqrt{\frac{2}{5}}\tau$, $q_1 = \frac{1}{2}(l_1 - l_2 + l_3 + l_4 + l_5)$ with radius $\sqrt{\frac{2}{5}}$ and $t_1 = \frac{1}{2}(l_1 + l_2 - l_3 + l_4 + l_5)$ with radius $\sqrt{\frac{2}{5}}\tau^{-1}$ where $\tau = \frac{1+\sqrt{5}}{2}$.

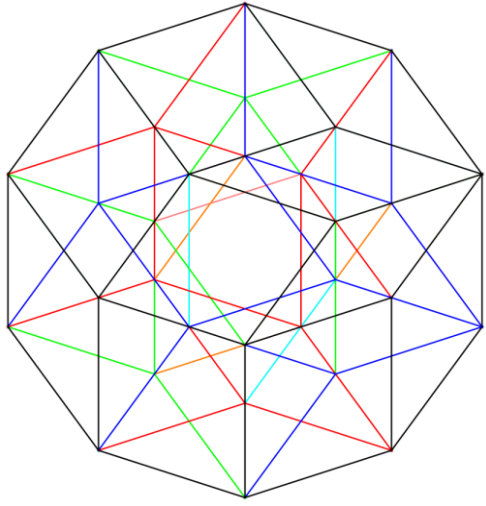

**Figure 13**
Projection of the Voronoi cell $V(0)$ of the lattice $\mathbb{Z}^5$ (edges connected to the origin are removed) which can be dissociated into ten decagons tiled by rhombs similar to those in Fig. 9. One of the decagons is depicted in Fig. 14.

The projected Voronoi cell $V(0)$, invariant under the dihedral group $W(I_2(10))$, consists of 10 overlapping decagons tiled by rhombs one of which is depicted in Fig. 14. The other 9 decagons tiled by Penrose rhombs are obtained by rotating the one in Fig. 14 by $\frac{\pi}{5}$ and its integer multiples.

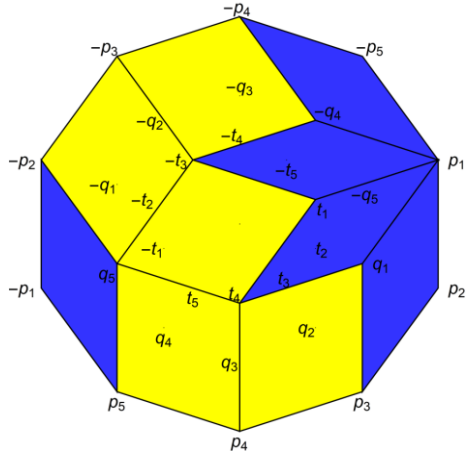

**Figure 14**
One version of tiling of the projected Voronoi cell $V(0)$ of $W_a(B_5)$ symmetric with respect to the line connecting the points $p_1$ and $-p_1$. The other octagons tiled by rhombs are symmetric with respect to the line segments connecting the pair of points $p_i$ and $-p_i$, $i = 2,3, \ldots, 10$.

Reflection elements of the dihedral group $W(I_2(10))$ are obtained from the reflection elements $R_1 := r_{l_2-l_3} r_{l_1-l_4} r_{-l_5}$ and $R_2 := r_{l_3-l_1} r_{l_4+l_5}$ by permutations of the vectors $l_i$. The symmetry of Fig. 14 is the diagonal line segment from $p_1$ to $-p_1$ through $\pm q_5$ and $\pm t_1$ and represented by the reflection element $r_{l_2-l_4} r_{l_1-l_5}$. Each of ten rotated copies of the Fig. 14 is symmetric with respect to the five line-segments represented by reflections obtained by permuting the vectors $l_i$ in $r_{l_2-l_4} r_{l_1-l_5}$. Fig. 14 is not symmetric with respect to the reflection elements such as $R_1 := r_{l_2-l_3} r_{l_1-l_4} r_{-l_5}$ which represents the line segment bisecting some edges of the concentric decagons.

A general set of generators of the affine dihedral group $W_a(I_2(10))$ can be written as

$$r_{l_2-l_3,n_2}r_{l_1-l_4,n_1}r_{l_5,n_5}, \quad r_{l_3-l_1,n_3}r_{l_4+l_5,n_4}, n_i \in \mathbb{Z}. \tag{31}$$

They leave the vector, $\lambda$, invariant:

$$\lambda = (n_1 + n_4 + \tfrac{n_5}{2})l_1 + \left(n_1 + n_2 + n_3 + n_4 + \tfrac{n_5}{2}\right)l_2 + \left(n_1 + n_3 + n_4 + \tfrac{n_5}{2}\right)l_3 + (n_4 + \tfrac{n_5}{2})l_4 + \tfrac{n_5}{2}l_5. \tag{32}$$

It is clear from (32) that if $n_i = 0$ the origin is invariant under the dihedral group.
One of the affine reflection elements of the affine dihedral group $W_a(I_2(10))$ is $r_{l_2-l_3}r_{l_1-l_4}r_{l_5,1}$ which leaves the line segment invariant from $p_1$ to $p_5$ through $t_2$ and $t_4$ in Fig. 14. The two reflections $r_{l_2-l_3}r_{l_1-l_4}r_{l_5,1}$ and $r_{l_2-l_4}r_{l_1-l_5}$ generate the dihedral group fixing the vertex $p_1$ invariant and the result is shown in Fig. 15 which represents a 10-fold symmetric tiling with thick and thin rhombs. We know that the weight lattice $\mathbb{Z}^5$ is tessellated by the Voronoi cell of the root lattice facet to facet that means there are joint cubes and squares leading to overlapping rhombs after projection onto the Coxeter plane. For example, the thin rhombus represented by vertices $p_5$, $q_5$, $-p_2$ and $-p_1$ can be matched with the appropriate rhombuses by translation and/or rotation of the tile in Fig. 14. By the elements of the affine dihedral group $W_a(I_2(10))$ we obtain the tiling in Fig. 16 by translating and rotating the decagon in Fig. 14 to fill the next layer of Fig.15 so that the finite patch is invariant under the finite group $W(I_2(10))$. This process continues to increase the number of tiles in the symmetric patch.

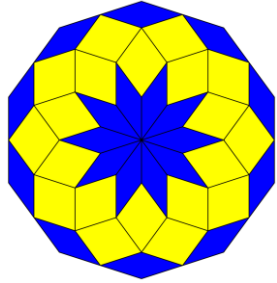

**Figure 15**
10-fold symmetric tiling by thin and thick rhombs obtained by rotation of Fig.14 around point $p_1$.

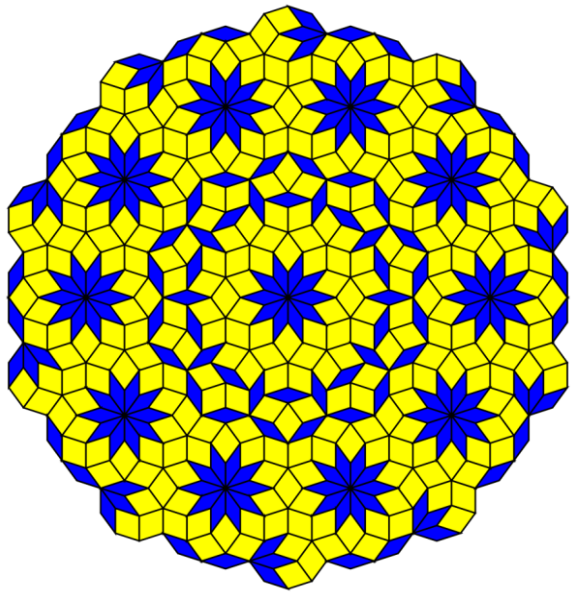

**Figure 16**
A patch of 10-fold symmetric tessellation of the Coxeter plane with Penrose rhombs by applying $W_a(I_2(10))$ on Fig.14.

## 6. Conclusion

In Section 2 we have introduced a general framework for the projection of the lattice $\mathbb{Z}^n$ and the tessellation of the Coxeter plane with rhombs under the affine dihedral subgroup $W_a(I_2(h))$ of the affine group $W_a(B_n)$. We have discussed the trivial cases for $n = 1,2$. The case $n = 1$ represents the infinite group generated by one reflection with respect to the origin and the affine reflection with respect to the point $\frac{1}{2}$ leading to the symmetry of the lattice $(m_1 + \frac{1}{2})l_1$. The group $W_a(B_2) \approx W_a(I_2(4))$ itself is the dihedral group representing the symmetry of the square lattice $\mathbb{Z}^2$. The only non-trivial case arises with $n = 3$ and the affine group $W_a(I_2(6))$ describes the projected hexagonal lattice as the projection of the simple cubic lattice.

Our main emphasis was on the projections of the lattices $\mathbb{Z}^n (n \geq 4)$ and the quasicrystallographic tiling scheme obtained by rhombs and the role of the symmetry of the affine group $W_a(I_2(2n))$ with $(n \geq 4)$. After giving the general procedure in Section 2. we have exemplified two cases with $n = 4$ and $n = 5$. We have shown that the Coxeter plane can be tiled with rhombs displaying local $2n$-fold symmetries. This technique can be extended to the projections of the lattices $\mathbb{Z}^6$ and $\mathbb{Z}^9$ leading to 12-fold and 18-fold symmetric tilings which can explain the corresponding quasicrystallographic structures. The technique we introduced emphasizes the role of the affine dihedral subgroup of the affine symmetry of the higher dimensional lattice in the tiling of the Coxeter plane and leads to the centrally symmetric tilings by rhombs.